\documentstyle[aps,12pt,epsf]{revtex}
\begin{document}

\title{PARTICLE REMOVAL REACTIONS WITH DEFORMED PROJECTILES}
\author{Alexander Sakharuk$^{1}$, P. Gregers Hansen$^{1,2}$, and 
Vladimir Zelevinsky$^{1,2}$}
\address{$^{1}$National Superconducting Cyclotron Laboratory and\\
$^{2}$Department of Physics and Astronomy,\\
 Michigan State University, East Lansing, MI 48824-1321 USA}
\date{\today}
\maketitle

\begin{abstract}
Single-particle removal reactions are becoming an important tool for
studying radioactive nuclei. The nuclei far from stability may reveal
new regions of large deformation. We discuss the influence of the
projectile deformation on  the stripping cross sections and
longitudinal momentum distributions of the core. The Glauber formalism
for describing such reactions is developed and compared with the
simpler geometric approach and with  existing experimental data. The
most significant effects  are expected for nuclei with valence
spherical orbitals strongly mixed by deformation. The shape of the momentum
distribution, in turn, can be used for determining the degree of deformation of
the projectile.
\end{abstract}

\section{INTRODUCTION}

Particle removal reactions of various types (nucleon transfer, knockout,
stripping and so on) have for a long time been one of the most powerful
tools for investigation of nuclear structure \cite{Butler,Sit:90}. 
We are here concerned with a class of reactions at high energies 
(``stripping" in the sense of the word used originally by Serber 
\cite{Ser:47}) of which the deuteron breakup is the archetype. 
The strong current interest in these reactions is that they, in inverse
kinematics, provide a unique opportunity
to obtain simultaneously partial probabilities for populating
individual final states leading to spectroscopic factors 
as well as momentum distributions characteristic of the quantum numbers of the
removed nucleon. This allows one to access detailed
information on the internal structure of projectile and residue. 

Nuclear breakup reactions are widely used nowadays for studying radioactive
nuclei \cite{Tani:95,Bar:96,Han:99,Tos:99}. A new method was suggested
recently for the spectroscopy with radioactive beams \cite{Nav:98}
which is well suited both for ``normal" and for halo nuclei. The
method uses the longitudinal momentum distribution of the projectile
residue (``core") for the determination of the orbital angular
momentum of the stripped nucleon. The identification is possible via
the comparison of the experimental distribution with the theoretical
one calculated, for instance, in the framework of the Glauber model
\cite{Han:96,Hen:96}. The association of the definite orbital momentum
value with  the shape of the longitudinal momentum distribution is
clearly justified for spherical nuclei. First, for a spherically symmetric
unpolarized projectile, the role of the spin-orbit interaction is not
essential for the stripping mechanism. Second, the longitudinal
momentum distribution is sensitive to the projection $m$ of the
orbital momentum of the stripped particle onto the beam direction. As
a rule, the largest of the possible $m$-values for a given $l$ value
dominates, and this $l$- and $m$-dependence allows one to determine
the orbital momentum more or 
less uniquely \cite{Bar:98}. The results of a recent series of
experiments with spherical or weakly deformed nuclei \cite{Nav:98}
confirm the applicability and flexibility of the method.

Recently the first attempt has been made to apply the same
techniques to strongly deformed projectiles. The inclusive reaction
$^{9}$Be($^{25}$Al,$^{24}$Mg)X was investigated in \cite{Nav:98}. Many
neutron-rich nuclei, for example neon and magnesium isotopes in the
region $N\approx 30$ (the island of inversion \cite{Wild:80}), are
expected to be strongly deformed, and the role of intruder
configurations, such as the $[330 \,1/2]$ level coming from the
$pf$-shell, is extremely important \cite{Hey:91}. This follows from the 
theoretical calculations \cite{Cau:98,War:90} and from the results
of the latest measurements of the transition probabilities
$B(E2;0^+_{g.s.}\rightarrow2^+)$ in neutron-rich $^{26,28}$Ne and 
$^{30,32,34}$Mg isotopes via intermediate-energy Coulomb excitation
\cite{Pri:99}. Available information on the structure of the nuclei in the 
inversion island is scanty, and one can hope to extract the shape
parameters from the analysis of the stripping reactions and 
corresponding momentum distributions.

However, in order to correctly interpret experimental data for the deformed
nuclei, one needs to modify essentially existing theoretical
approaches. Several features of nuclear deformation should 
be taken into account consistently in the reaction description:
(i) the deformed mean field of the core is responsible for the strong
mixing of different orbital momenta of the stripped valence nucleon;
(ii) although in the case of axial symmetry the projection of the
nucleon total angular momentum on the symmetry axis 
is still a good quantum number (usually denoted as $\Omega$), 
the energy splitting and therefore difference
in the population of different doubly degenerate $(\pm \Omega)$ 
levels preclude complete averaging 
and stress the effects of the spin-orbit coupling; 
(iii) pure intrinsic $\Omega$-states
of the nucleon have no definite angular momentum projection in the laboratory 
reference frame defined by the beam direction. 
All mentioned elements are well known but we
have to incorporate them consistently in the reaction formalism.
The realistic calculations along these lines imply cumbersome and
time consuming numerical work. This might be one of the reasons why,
in spite of the growing interest in reactions with deformed nuclei
\cite{Rid:98,Chr:98}, our attempt is apparently the first one of this
kind. 

Of course, scattering by nonspherical nuclei has been studied
long ago. The most general formalism for this purpose is the coupled
channel method outlined in the book \cite{Boh:74}. Unfortunately, it is 
rather hard to use this technically complicated method 
for the description of realistic reactions. A considerably simpler
approach, namely the adiabatic approximation, was proposed by Drozdov
and Inopin in the middle of the fifties \cite{Dro1:55,Ino:56}. They
used the partial wave  expansion of the scattering wave function
and assumed that the collective motion (rotation or vibration) of the
deformed target nucleus is frozen during the collision time, being
very slow compared to other degrees of freedom. The term ``adiabatic''
refers to the orientational and shape degrees of freedom only and
should not be confused with the adiabaticity with respect to valence
nucleon motion. The latter assumption is also frequently
used in Glauber theory for the description of the reaction mechanism.
The Glauber adiabaticity means that, starting at energies of 50-100
MeV per nucleon, the characteristic time scales for the relative
projectile-target motion and for the nucleon internal motion inside
the projectile are getting very different.

In the framework of the orientational adiabatic approximation, it is
impossible to examine the deep excitations of the core. At the same
time, soft excitations, with no essential reorganization of the
internal core structure,  are quite treatable. Typical energies of
collective rotational states are low \cite{Boh:74} compared to
characteristic reaction energies. In some sense we have here nearly
the same situation as in perturbation theory for close or degenerate
states. To be consistent with the adiabatic approximation, we 
fix the orientation of interacting objects during the
collision. The state with a certain orientation is a wave packet of
stationary angular momentum states. This means that after the
collision various rotational members of the superposition can be
excited. 

At the end of the seventies, Abgrall {\sl et al.} \cite{Abg:79}
combined the coupled channel formalism with the Glauber scattering
theory to describe the excitation of the ground state rotational band
by medium energy (1 GeV) protons. The regular treatment of the
adiabatic approximation in the framework of the Glauber theory was
given in 1990 by F\"{a}ldt and Glauber \cite{Fal:90}. They treated
only soft rotational excitations without restructuring of the internal
projectile wave function. The deformation effect on the stripping
cross section was studied only for the relativistic deuteron stripping
\cite{Fal:80} (see also \cite{Nis:82} for a qualitative discussion
of the deformation effect on the tensor analyzing power for light
projectiles). The approach of our work, being a natural
continuation of the guideline \cite{Fal:90}, is, on the other hand,
close to the formulation suggested  in the recent paper by Christley
and Tostevin \cite{Chr:98}. These authors came to a conclusion that
the deformation is not very important for the total (angle-integrated)
reaction cross sections: the corresponding change is
less than 1$\%$. Our results are consistent with this
conclusion. However, the longitudinal momentum distributions 
and partial cross sections for
stripping from deformed orbitals, which are a strong mixture of
spherical single-particle states, are sensitive to deformation.

The paper is organized as follows. The semi-microscopic reaction
formalism, taking into account the deformation of the projectile, is
developed in Sec.~\ref{se1}; the influence of the deformation on the
stripping reaction is analyzed, and expressions for the cross
section and the longitudinal momentum distribution of the projectile
residue are obtained. Sec.~\ref{se2} is devoted to the study of the
deformation effects on the core longitudinal momentum distribution in the
inclusive, with respect to the rotational states of the core,
formulation. This study is performed using a geometric approach 
\cite{Han:96} which is simpler and more transparent than the rigorous
treatment of Sec.~\ref{se1}. In Sec.~\ref{se3} numerical calculations
of the stripping cross sections  and longitudinal momentum
distributions are presented and discussed in the framework of the
approaches of Sec.~\ref{se1} and Sec.~\ref{se2} for
different final states of the residual nucleus in a typical reaction 
$^{9}$Be($^{25}$Al,$^{24}$Mg)X. The interplay of the internal
structure of the projectile and the reaction mechanism is illustrated for
the model $^{9}$Be($^{28}$P,$^{27}$Si)X reaction. We discuss 
the perspectives for further studies and propose
new experiments in a possible region of large deformation
in neutron-rich Mg isotopes. Summary and 
final remarks are made in the concluding section.
 
\section{STRIPPING CROSS SECTION FOR A DEFORMED PROJECTILE}
\label{se1}

In the description of projectile breakup reactions, the established
terminology is slightly different from that conventionally used in the
theory of direct reactions \cite{Butler}. Two processes are usually
distinguished, stripping (or absorption) and diffraction (or elastic
breakup). In this word usage (see, for example, \cite{Yab:92,Fuj:80}),
elastic breakup is the process in which the projectile core and the
valence, in particular halo, particle are in the continuum states
after the interaction with the target. In addition, the projectile
residue (core) and the target remain in their respective ground
states. We generalize this definition to include possible ``soft"
excitations of the core nucleus, such as collective rotations, which
disturb the intrinsic structure in a minimal way. The absorption
differs from the elastic breakup by possible excitations of the core
and disappearance of the valence particle in the final state due to
the absorption by the target. In this work we restrict overselves by
the treatment of the stripping/absorption type of breakup reactions
only. This is, generally, the dominant channel \cite{Tos:99}, except for the halo
states. Before considering the reaction in detail, let us agree on a
general framework of the approach. 

Target and core nuclei are treated as structureless particles
in the actual calculations. All the observables depend on the internal
structure of the core and the target only indirectly via corresponding
interaction potentials and density functions. Nevertheless, we retain
the explicit dependence on the internal Jacobi coordinates of the
colliding nuclei in the derivation of the basic formulas. This makes
the derivation more transparent and shows the way for extending the
approach. In addition, the core may be deformed, in contrast to
previous treatments \cite{Yab:92,AlK:95,Hen:96}. Therefore we need to
fix its orientation (more precisely, the orientation of the core mean
field which determines the geometry of the orbitals for the valence
nucleon). Although we assume here that the core is axially symmetric,
the approach can be generalized for triaxial deformation that would be
more suitable for nuclei such as, for example, $^{24}$Mg. In a similar
way, the deformation of the target can be accounted for as would be
necessary for transfer reactions. 

The interaction of a valence nucleon (for definiteness, a proton)
with the target, p+T, and the core-target, C+T, interaction are
described in terms of phenomenological optical potentials. For the C+T  
interaction, the convolution of a nucleon-target phenomenological
potential with the core density is used. We do not employ the
traditional Glauber approach based on  the nucleon-nucleon scattering
amplitudes \cite{Gla:59,Gla:70} because of relatively small kinetic
energy per nucleon in the projectile in the experiments we try to
describe. It is known, however, that the region of validity of the
Glauber approximation extends considerably beyond what one could
expect from the approximations made in the standard derivation
\cite{Hen:96}. Therefore, our approach can turn out to be very similar
to that of Glauber.       

\begin{figure}
\centerline{
\epsfxsize=10.0cm \epsfbox{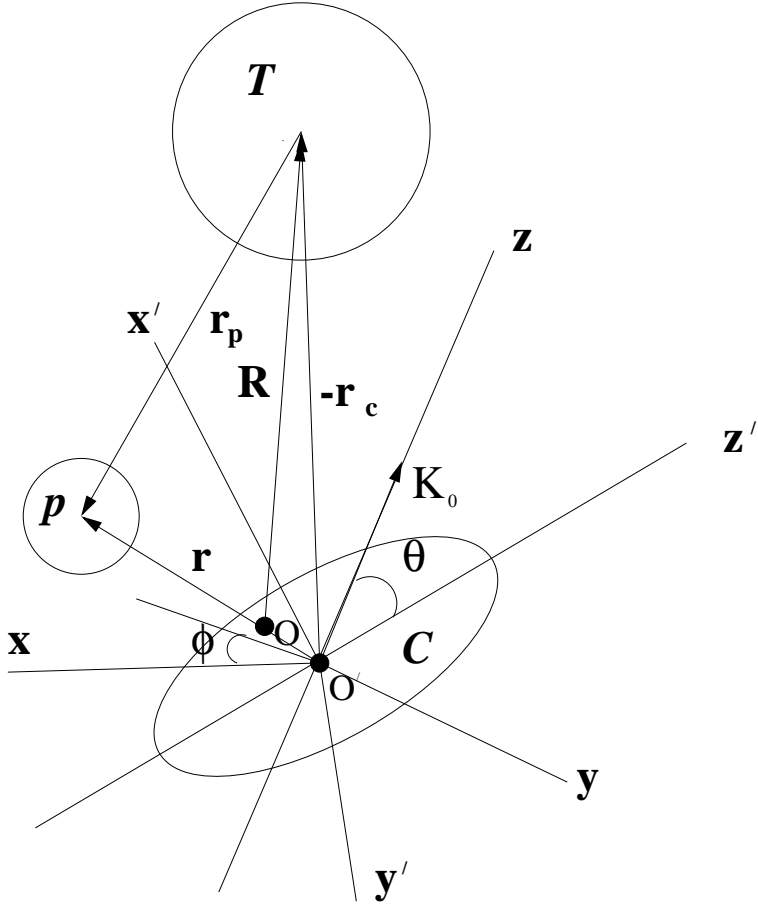}}
\caption{Coordinates used for the description of the reaction.}
\label{fff0}
\end{figure}

The calculations are performed either in the ``laboratory'' frame, 
with the origin coinciding with the projectile center-of-mass $O$ and
$z$-axis along the beam direction, or in the body-fixed frame,
characterized by the origin at the core $O'$ and $z'$-axis along the
core symmetry axis. The positions of the target in these frames are
given by the vectors ${\bf R}$ and $-{\bf r}_{c}$, respectively. 
The set of coordinates used in our calculations is shown in
Fig.~\ref{fff0}. 

The description of stripping reactions with exotic nuclei has its
roots in the early works on the deuteron stripping 
\cite{Ser:47,Sit:57}. Attempts to construct a more regular theory of
stripping processes have appeared considerably later
\cite{Fuj:80,Hus:85,Hus:89}. Referring to the deuteron stripping
reaction formalism as a suitable framework for exotic nuclei
fragmentation \cite{Ban:93}, it is assumed by analogy that a stripped
projectile constituent (proton) interacts strongly with the target nucleus. 
The core of the projectile escapes as an analog of the neutron. The only
difference is the possibility for the core to be excited during the
collision. The total stripping cross section in such an approach is
analogous to the proton stripping from the deuteron \cite{Sit:90}. Below we
sketch briefly the formal derivation of the total stripping cross
section and the core longitudinal momentum distribution.

The projectile (core + valence nucleon) initial 
wave function is  
\begin{equation}
|\chi_i\rangle = e^{-i{\bf K}_0\cdot{\bf R}}
|\phi_i\rangle,
\label{loqe}
\end{equation}
where the plane wave describes motion of the center-of-mass with respect to the
target, and the initial intrinsic wave function of the projectile 
\begin{equation}
|\phi_i\rangle = \sum_{\nu\alpha} {\cal A}^{(\alpha)}_{\nu} a^{\dagger}_{\nu}
\Phi^{(\alpha)\dagger}_{\nu}|0\rangle
\label{swd}
\end{equation}
is a sum of antisymmetrized products of the core internal wave functions
generated by the operator $\Phi^{(\alpha)\dagger}_{\nu}$ and the 
function of the relative
proton-core motion (creation operator $a^{\dagger}_{\nu}$). Here the superscript
$\alpha$ enumerates various core states, $|\nu\rangle$ 
are single-particle orbitals, and ${\cal A}^{(\alpha)}_{\nu}$
are corresponding spectroscopic amplitudes. The calculation of the
spectroscopic amplitudes for all possible combinations $(\nu\alpha)$ 
requires very specific and hard work. Frequently only one possibility
is allowed or is taken into account, which corresponds to a definite
configuration, for example to the ground state 
of the spherical (deformed) core + the proton occupying a certain spherical 
(deformed) orbital. This approximation works fairly well in the case
of light or halo projectiles when there are no bound excited
states. The situation is more complicated for projectile
nuclei which possess many particle-bound states. We limit ourselves by using
the same approximation with the spectroscopic amplitudes calculated in
the framework of the particle-rotor model, consistent with our general
approach, or borrowed from the shell-model calculations
\cite{War:92}. In the case of a deformed core, the internal wave
function contains information about the orientation of the core
symmetry axis with respect to the space-fixed frame (the corresponding
spherical harmonic $Y_{L_iM_{L_i}}(\Omega_a)$), and the
single-particle quantum states $|\nu\rangle$ are quantized onto the
intrinsic symmetry axis. 

The situation with the final state wave function is even
more complicated because here we have to sum over all possible final 
configurations of the core and valence proton, including the ones
where the knocked out proton is absorbed by the target. We do not
consider the exact structure of this wave function. Instead we
determine \cite{Sit:90,Sit:57} the absorption cross section through 
a deficit of the probability for the proton to be found at some
point ${\bf r}_p$ that appears as a result of the p+T interaction. 
Such an approach is based
essentially on the spectator mechanism of the stripping reaction and
is adequate only in a narrow kinematic region. This mechanism
presupposes that the proton is stripped off the projectile while the
remaining core fragment continues its motion essentially without being
disturbed except for a smooth action of the optical core-target potential
$V_{{\rm C+T}}$ taken into account by the eikonal core phase shift.
The incident energy of the projectile has to be sufficiently
large and the scattering angle of the core must be small to ensure the
applicability of the eikonal approximation \cite{Fuj:80,Hus:85}.
It is assumed in the framework of the spectator mechanism that the
nucleus-nucleus interaction time is too short for a change of the core
symmetry axis direction. As we mentioned, this means 
the possibility to reveal various rotational components of the frozen
orientation. Therefore the final-state deformed core wave function
$\Phi^{(\alpha)\dagger}_{\nu}|0\rangle$ contains a spherical harmonic
$Y_{L_fM_{L_f}}(\Omega_a)$, which defines the same direction of the
core symmetry axis as before the projectile breakup (adiabatic
approximation), but in general with $L_{f},M_{f}\neq L_{i},M_{i}$. 

According to the idea of the Glauber approximation and our hypotheses of the
reaction mechanism, the reaction yield
is determined by the profile functions for the 
propagation of the valence
proton, $S_{p}$, and the core, $S_{c}$, in the field of the target potentials,
p+T and C+T, respectively \cite{Yab:92},
\begin{equation}
S_p({\bf b}_p,z_p) = \exp\left(-\frac {i}{v}\int_{-\infty}^{z_p} 
V_{{\rm p+T}}({\bf b}_p,z'_p)dz'_p\right), 
\end{equation}
\begin{equation}
S_c({\bf b}_c,z_c) =\exp\left(-\frac {i}{v}\int_{-\infty}^{z_c} 
V_{{\rm C+T}}({\bf b}_c,z'_c)dz'_c\right).
\label{tcsr3}
\end{equation}
Here and below we have adopted units with $\hbar=1$. The coordinates
${\bf r}_p$ and ${\bf r}_c$ are expressed from Fig.~\ref{fff0} as
\[ {\bf r}_p = \frac {A_c}{A_c+1}{\bf r}  - {\bf R}, \quad
{\bf r}_c = - \frac {1}{A_c+1}{\bf r} - {\bf R}. \] 
The impact parameter vectors
${\bf b}_p$ and ${\bf b}_c$ are their
corresponding two-dimensional projections onto the plane perpendicular
to the beam direction.

In the same spirit, the (core + proton) Glauber wave function
in the final state, projected onto definite states of core and proton 
motion with momenta ${\bf k}_c$ and ${\bf k}_p$, respectively, is
\begin{equation}
|\chi_f^{(-)}\rangle =  \tilde{S}_{p}({\bf b}_{p},z_{p})
\tilde{S}_{c}({\bf b}_{c},z_{c})e^{i{\bf k}_c\cdot{\bf r}_c}
e^{i{\bf k}_p\cdot{\bf r}_p}\Phi^{(\alpha)\dagger}_{\nu}|0\rangle,
\label{cvyu}
\end{equation}
where 
\begin{equation}
\tilde{S}_{p}({\bf b}_{p},z_{p})=\exp\left(\frac {i}{v}\int^{\infty}_{z_p}
V_{{\rm p+T}}({\bf b}_p,z'_p)dz'_p\right),                     \label{sp}
\end{equation}
and
\begin{equation}
\tilde{S}_{c}({\bf b}_{c},z_{c})=\exp\left(\frac {i}{v}\int^{\infty}_{z_c}
V_{{\rm C+T}}({\bf b}_c,z'_c)dz'_c\right).                  \label{sc}
\end{equation}
Here we again take into account the interactions with the target
by means of the eikonal phase shifts \cite{Hus:85}. 

The stripping reaction is defined essentially as disappearance of the proton
from the projectile as a result of the collision with the target. We need to
compare the probabilities for the proton to be found at some point 
${\bf r}_{p}$ with and without the p+T interaction. The probability amplitude 
for the core in a given configuration $\alpha$ to have the momentum
${\bf k}_c$ and for the valence proton, emerging from the intrinsic state 
$|\nu\rangle$ with the spectroscopic amplitude ${\cal A}_{\nu}^{(\alpha)}$,
to be found at the position ${\bf r}_p$ is 
\begin{equation}
a({\bf k}_c,{\bf r}_p) = {\cal A}_{\nu} S_{p}e^{-i{\bf k}_p\cdot{\bf r}_p}
\int d^3r_c\, e^{-i{\bf k}_c\cdot{\bf r}_c + i{\bf K}_0\cdot{\bf R}} 
\phi_{\nu}({\bf r}) \langle 0|\Phi_{\nu}S_c 
\Phi^{\dagger}_{\nu}|0\rangle,
\label{dfrte}
\end{equation}
where ${\bf r}={\bf r}_{p}-{\bf r}_{c}$, 
$\phi_{\nu}({\bf r})=\langle {\bf r}|a^{\dagger}_{\nu}|0\rangle$
is the function of the relative proton-core motion in the projectile
nucleus, and we omit the arguments of the profile functions
as well as the superscript $\alpha$. The phase integrals in the 
profile functions $S_c$ and $S_p$ in (\ref{dfrte}) are now taken from 
$-\infty$ to $\infty$ and do not depend on
the $z$-components of the core and proton radius-vectors. 
The intrinsic matrix element in (\ref{dfrte}) and the profile
function $S_c$ contain the orientation parameters of the deformed core
which are adiabatically fixed during the collision. Therefore the whole
amplitude $a({\bf k}_{c},{\bf r}_{p})$ is still an operator in orientation
angles. Thus, apart from the stripping probability, it will determine the
rotational excitations of the remaining core. 

The core final momentum is a sum of $(A_c/A){\bf K}_0$ (the core part
of the initial momentum of the projectile as a whole) and the internal
momentum ${\bf q}$ corresponding to relative core-proton motion inside
the projectile. Due to the large mass difference, $A_c\gg A_p$, we can 
put $A_c\approx A$ and obtain for the exponent in the integrand (\ref{dfrte})
\[ -i{\bf k}_c\cdot{\bf r}_c - i{\bf K}_0\cdot{\bf R} =
-i{\bf q}\cdot{\bf r}_c. \]
Eq.~(\ref{dfrte}) thus can be rewritten as
\begin{equation}
a({\bf k}_c,{\bf r}_p) = {\cal A}_{\nu} S_{p} e^{-i{\bf k}_p\cdot{\bf r}_p}
\int d^3r_c e^{-i{\bf q}\cdot{\bf r}_c} 
\phi_{\nu}({\bf r}) \langle 0|\Phi_{\nu}S_c 
\Phi^{\dagger}_{\nu}|0\rangle,
\label{dfrt}
\end{equation}
Without the p+T interaction, the expression (\ref{dfrt}) reduces to
\begin{equation}
a_{0}({\bf k}_c,{\bf r}_p) = {\cal A}_{\nu} e^{-i{\bf k}_p\cdot{\bf r}_p}
\int d^3 r_c \,e^{-i{\bf q}\cdot{\bf r}_c} 
\phi_{\nu}({\bf r}) \langle 0|\Phi_{\nu}S_c
\Phi^{\dagger}_{\nu}|0\rangle.
\label{dftk}
\end{equation}
The probability for the proton to get absorbed in different spatial
points ${\bf r}_p$, provided that the core escapes with the momentum
${\bf k}_c$, is proportional to the difference of the squares of the
absolute values of the orientational matrix elements of the operators
(\ref{dftk}) and (\ref{dfrt}) between the given rotational states of the core,
\[ |a_{0}({\bf k}_c,{\bf r}_p)|^{2}_{fi} - |a({\bf k}_c,{\bf r}_p)|^{2}_{fi} = 
|{\cal A}_{\nu}|^2 (1-|S_p({\bf b}_p)|^2) \]
\begin{equation}
\times\left|\int d^3 r_c\, e^{-i{\bf q}\cdot{\bf r}_c}\int d\Omega_{a}\, 
\phi_{\nu}({\bf r}) Y^*_{L_fM_{L_f}}(\Omega_a)S_c({\bf b}_c)
 Y_{L_iM_{L_i}}(\Omega_a) \right|^2.
\label{drtm}
\end{equation}
Here we have explicitly written the arguments of the profile
functions and preserved only those parts of the core wave
functions that define the orientation of the core mean field. In the
integrand we have the product of four factors: two spherical harmonics
defining the core rotational state before and after the interaction
with the target, the core profile function (our operator), and the
function of the proton-core relative motion $\phi_{\nu}({\bf r})$. The
dependence of $\phi_{\nu}({\bf r})$ upon the core orientation is
considered in detail in the next section. 

The approach above ensures the correct
limit of spherical symmetry. Indeed, in this case all factors in
(\ref{drtm}), except the spherical functions, do not depend on the
orientation angles $\Omega_a$, and the orthonormalization of the
spherical harmonics singles out the diagonal transition of the
core. Another way to consider the core orientation would be to retain the 
spherical harmonics up to the final calculation of the stripping
differential/total cross section and then average over all possible
directions of the core symmetry axis. This would however imply
ignoring interference contributions between the processes of
rotational recoil of the core occurring at different positions of the
core. In addition, in that case we encounter a technical problem of
summation over all possible final  projections $M_{L_f}$ (final
channels). Since the values of  $M_{L_f}$  are connected with the
orientation of the core axis $\Omega_a$, the simple summation can lead
to the double counting.

The differential cross section for the core final
momentum $d^3k_c=d^3q$ is obtained, as in the deuteron stripping theory 
\cite{Sit:90}, after integrating the difference of probabilities (\ref{drtm}) 
on the interaction plane, orthogonal to the projectile direction, over the 
proton impact parameter ${\bf b}_p$, 
\[ \frac {d\sigma}{d^3q} = \frac {1}{(2\pi)^3} \overline{\sum} 
|{\cal A}_{\nu}|^2 \int d^2b_p\left(1-\left|S_p({\bf b}_p)\right|^2\right) \]
\begin{equation}
\times \left| \int d^3 r \,e^{-i{\bf q}\cdot{\bf r}}
\int d\Omega_a \,\phi_{\nu}({\bf r})
Y^*_{L_fM_{L_f}}(\Omega_a)
S_c({\bf b}_c) Y_{L_iM_{L_i}}(\Omega_a)\right|^2, 
\label{jiopp}
\end{equation}
where $\overline{\sum}$ means, as usual, the averaging over the
initial angular momentum projections and summation over the final
ones. Here we made a change of variable from the set 
${\bf r}_p$ and ${\bf r}_c$ to the set ${\bf r}_p$ and ${\bf r}={\bf
r}_c-{\bf r}_p$. The three-dimensional momentum distribution of the core
(\ref{jiopp}) differs from the results of refs.~\cite{Yab:92,Hen:96}
by the presence of the spherical harmonics with the additional
integration over $\Omega_a$.   

The longitudinal momentum distribution of the core is obtained 
by integrating over the transverse components ${\bf q}_{\perp}$,
\[ \frac {d\sigma}{dq_z} = \frac {1}{2\pi} \overline{\sum} 
|{\cal A}_{\nu}|^2 \int d^2b_p\left(1-\left|S_p({\bf b}_p)\right|^2\right) \]
\begin{equation}
\times  \int dxdy   \left| \int dz e^{-iq_{z}z}
\int d\Omega_a \phi_{\nu}({\bf r})
Y^*_{L_fM_{L_f}}(\Omega_a)
S_c({\bf b}_c) Y_{L_iM_{L_i}}(\Omega_a)\right|^2. 
\label{jiop5}
\end{equation}
The total proton stripping cross section follows after
the integration over $q_z$, 
\begin{equation}
\sigma = \overline{\sum} |{\cal A}_{\nu}|^2 
\int d^{2}b_{p}\,\left(1-|S_{p}({\bf b}_{p})|^{2}\right)\int d^3r \left| \int
d\Omega_a \phi_{\nu}({\bf r}) Y^*_{L_fM_{L_f}}(\Omega_a)
S_c({\bf b}_c) Y_{L_iM_{L_i}}(\Omega_a)\right|^2. 
\label{polop}
\end{equation}
The expression obtained in a number of works \cite{Yab:92,Hen:96}
differs by the absence of the angular integral. In fact, these results 
are known from the theory of deuteron stripping
\cite{Sit:90,Sit:57}, of course again without deformation effects.
The more complicated and formal approaches, such as, for example,
based on Ref.~\cite{Fuj:80}, do not bring any real advantage. 

The presence of the angular integral in (\ref{jiop5},\ref{polop}),
together with the angular dependence of the core profile function
$S_c({\bf b}_c)$ and, especially, the proton wave function
$\phi_{\nu}({\bf r})$, makes a practical calculation of the
longitudinal momentum distribution (\ref{jiop5}) and the total
stripping cross section (\ref{polop}) a difficult computational problem. Eq.
(\ref{polop}), for example, contains seven embedded integrals. In
addition, the function $\phi_{\nu}({\bf r})$ and its arguments depend
on the angular variables in a very complicated way. To ensure the
resonable accuracy one needs to carefully check the convergence of each
numerical integration. In practice we have used a time consuming
combination of the grid integration with the Monte Carlo methods. 
We guess that these technical problems were and
remain up to now the main obstacle preventing the consistent account of the
deformation effects in the calculation of reaction observables. 

\section{``GEOMETRIC'' DESCRIPTION: LONGITUDINAL MOMENTUM DISTRIBUTION
OF THE CORE} 
\label{se2}

The longitudinal momentum distribution may be calculated directly from
(\ref{jiop5}). Technically, this is very time-consuming. 
Indeed, we need to perform a multidimensional integration over 
single-particle and core coordinates and over the orientational angles
$\Omega_a$. Prior to that, we have to calculate the core profile
function $S_{c}$ as a convolution of the nucleon-target potential with a core
density distribution which, in turn, strongly depends on the
orientation of the core symmetry axis. To avoid these technical
difficulties, one can use a simpler approach
developed in \cite{Han:96} on the ``physical" level of rigor. 
                                                            
In the approach \cite{Han:96} we assume that the profile functions are step
functions (``black disc" approximation).
Further, the interaction time is very short (a sudden
approximation) so that the core after reaction retains the same momentum
distribution as in the projectile before the collision
with a target. Due to conservation of the total momentum,
the core momentum distribution is determined by that for the
stripped valence nucleon inside the projectile. Therefore, one needs to
calculate the the three-dimentional Fourier transform 
of the part of the initial valence nucleon wave function which is
cut out by the target. Geometrically the overlapping region of the
valence nucleon wave function with the target corresponds
\cite{Han:96} to a cylinder with the axis along the beam direction and
the radius defined \cite{Tani:95} by the effective radius of the target
$R_T$. 

The approach \cite{Han:96} can be extended straightforwardly in order
to take into account the projectile deformation. It is easy to include 
the case of deformed targets also but with significant computational
complications. In what follows we consider nucleon stripping from a
deformed projectile on a spherical target. 

The core mean-field deformation can be treated in the framework of the 
standard Nilsson model in the projectile body-fixed
frame \cite{Boh:74,Nil:55,Nil:95}. Double degenerate single-particle
orbitals labeled by the asymptotic quantum numbers 
$[Nn_z\Lambda\Omega]$ are obtained by the
diagonalization of the phenomenological nuclear Hamiltonian 
in the basis of spherically symmetric
wave functions  
\begin{equation}
|[Nn_z\Lambda\Omega]\rangle = \sum_{lj}
 \alpha^{[Nn_z\Lambda]}_{lj\Omega} |lj\Omega\rangle.
\label{sdfgk}
\end{equation}
Note that the spin-orbit coupling is essential for the correct
definition of single-particle states. 
For the $sd$-shell nuclei we have six double degenerate
basis states $|lj\Omega\rangle$
some of which (with the same angular momentum projection $\Omega$ onto
the core symmetry axis) are coupled due to the static core
deformation. The corresponding single-particle orbitals in
the laboratory frame result from the rotation
$\hat{T}({\cal R})$, 
\begin{equation}
\psi({\bf r}) = \hat{T}({\cal R})|[Nn_z\Lambda\Omega]\rangle =
\sum_{lj} \alpha^{[Nn_z\Lambda\Omega]}_{lj\Omega}
\sum_{\Omega'=-j}^{j} D^j_{\Omega'\Omega}({\cal R}) |lj\Omega',\left\{
\hat{{\cal R}}{\bf r}\right\}\rangle,
\label{lkas}
\end{equation}
where $D^j_{\Omega'\Omega}({\cal R})$ are elements of the rotation matrix
\cite{Edm:60}.
The operator $\hat{{\cal R}}$ is defined by the
Euler angles $\theta_1,\theta_2,\theta_3$ but one of
these angles is arbitrary in the case of axial symmetry
and may be set equal to zero. The remaining
two are the polar and azimuthal angles between the core symmetry axis
and the beam direction, $\theta_1=\varphi$, $\theta_2=\vartheta$. 

The three-dimensional nucleon momentum distribution inside the
projectile is the squared Fourier transform of the wave function (\ref{lkas})
\begin{equation}
\tilde{\psi}({\bf q}) = \int d^3r
e^{-i{\bf q}\cdot{\bf r}}
\psi({\bf r}). 
\end{equation}
For a specific reaction, the momentum distribution corresponds to the 
spatial integration restricted by the region $G$ that, in the plane 
perpendicular to the beam ($z$) direction, 
is a disk with the center at the point ($x=b_{min},y=0,z=0$) and the
radius $R_T$. Here $b_{min}$ is the lowest possible value of the impact
parameter $b_{min}=R_c+R_T$, $R_c=r_oA_c^{1/3}$ is the effective core
radius (it is supposed that the interactions with the impact parameter
less than $b_{min}$ lead to violent collisions and cannot be seen in the
stripping channel). 
The longitudinal momentum distribution is obtained by
integrating $|\tilde{\psi}|^{2}$ 
over the transverse ${\bf q}$ components,
\begin{equation}
\frac {dW}{dq_z} = \int \left|\tilde{\psi}({\bf q}_{\perp},
{\bf q}_z)\right|^2 \frac{d^2q_{\perp}}{(2\pi)^{3}} = \frac {1}{2\pi}
\int_{G} dxdy \left| \int_{-\infty}^{\infty} dz\,e^{iq_z z}
\psi(x,y,z) \right|^2.
\end{equation} 
Using eq. (\ref{lkas}) we obtain
\[ \frac {dW}{dq_z} = \frac {1}{2\pi} \int_{G} dxdy \left|
\int_{-\infty}^{\infty} dz\,e^{iq_z\cdot z} \sum_{lj}
\alpha^{[Nn_z\Lambda]}_{lj\Omega}
\sum_{\Omega'=-j}^{j} D^j_{\Omega'\Omega}({\cal R})\right. \]
\begin{equation}
\left. \times \sum_{\Lambda} (l\Lambda,1/2\, \Omega'-\Lambda|j\Omega')
R_{nl}(r) Y_{l\Lambda}(\hat{{\bf r}})
\chi_{1/2\,\Omega'-\Lambda}  \right|^2.
\end{equation}
Taking into account the orthogonality of the spin functions
$\chi_{1/2\,\Omega'-\Lambda}$, after some algebra we obtain the final
expression for the longitudinal momentum distribution of the
stripped nucleon 
\[ \frac {dW}{dq_z} = \frac {1}{2\pi} \int_{G} dxdy \sum_{\sigma=\pm
1/2}  \left| \sum_{lj} \alpha^{[Nn_z\Lambda]}_{lj\Omega}
\sum_{\Omega'=-j}^{j} D^j_{\Omega'\Omega}({\cal R})
(l\,\Omega'+\sigma,1/2\,-\sigma|j\Omega') \right. \]
\begin{equation}
\left. \times\int_{-\infty}^{\infty}dz\,
e^{iq_zz}R_{Nl}(\hat{{\cal R}}r)Y_{l\,\Omega+\sigma}
(\hat{{\cal R}}\hat{{\bf r}}) \, \right|^2.
\label{frty}
\end{equation}
The single-nucleon stripping cross section is obtained by integrating
over $q_z$,
\[ \sigma =  \int_{G} dxdy \sum_{\sigma=\pm
1/2} \int_{-\infty}^{\infty}dz\, \left| \sum_{lj}
\alpha^{[Nn_z\Lambda]}_{lj\Omega} 
\sum_{\Omega'=-j}^{j} D^j_{\Omega'\Omega}({\cal R})
(l\,\Omega'+\sigma,1/2\,-\sigma|j\Omega') \right. \]
\begin{equation}
\left. \times
R_{Nl}(\hat{{\cal R}}r)Y_{l\,\Omega+\sigma}
(\hat{{\cal R}}\hat{{\bf r}}) \, \right|^2.
\label{frty2}
\end{equation}

The longitudinal momentum distribution (\ref{frty}) and total
stripping cross section (\ref{frty2}) are obtained at a fixed
direction of the core symmetry axis. We assumed that the core can
not change its orientation during the interaction time. But we did not
impose any other limitations on the core state of motion -
the core can be cranked by the process of interaction obtaining any
angular momentum allowed by selection rules. Since we do not
specify the state of the core exactly we have here an ``inclusive'' 
process with respect to the core final states. It means, in turn, that
the cross section (\ref{frty2}) is to be understood as a sum of partial
cross sections corresponding to the excitation of the particular core
rotational states and calculated in the framework of the more regular
approach of Sec.~\ref{se1} where all interference features were fully
accounted for. In distinction to that, in the geometric description
the stripping processes intitated at different positions of the core
are added incoherently and,
for the unpolarized projectile, we have to average the $dW/dq_z$,
eq.~(\ref{frty}), over all possible orientational angles  $\varphi$,
$\vartheta$. It gives an 
additional double integration and the factor $1/4\pi$ in
(\ref{frty},\ref{frty2}). We do not have any averaging in the 
approach of Sec.~\ref{se1} where the integration over the
orientational angles appears 
as a part of the calculation of the matrix element over the core
internal wave functions. An attempt to carry out the averaging leads to
uncertainties in the summation over core final states (double counting
mentioned in Sec.~\ref{se1}). 

\section{RESULTS AND DISCUSSION}
\label{se3}

For practical calculations we use the Nilsson Hamiltonian (with
constants from \cite{Boh:74}) expressed in the dimensionless stretched
coordinates \cite{Nil:95}. The coordinate dependence of the basis wave
functions in (\ref{lkas}) upon the rotation operator noticeably
lengthens the computer calculations. The harmonic oscillator wave
functions also have an incorrect asymptotic behavior, decaying too
fast at large distances. This shortcoming is not very serious 
for tightly bound projectiles (with the binding energies
$\geq$5 MeV). However for the description of weakly bound 
nuclei, the realistic behavior of single-particle wave functions may be crucial
so that in this case the results obtained with the aid of the Nilsson
Hamiltonian should be considered as preliminary estimates.

We start with the inclusive reaction $^{9}$Be($^{25}$Al,$^{24}$Mg)X
studied experimentally in \cite{Nav:98}. The deformed nature of the
$^{24}$Mg nucleus \cite{Boh:74} and, correspondingly, $^{25}$Al
projectile, is well established.
Our analysis based on the expression (\ref{frty}) can be considerably 
simplified in the case of the observation of the core nucleus in its 
ground state. Indeed, the valence proton in $^{25}$Al occupies the
Nilsson orbital $[202\,\,5/2]$. This orbital is generated by the
spherical $j=5/2, \, m_j=5/2$ basis function which is not 
mixed by the core deformed mean field with other spherical orbitals
from the $sd$-shell so that we have only a slight deformation
dependence through the use of the stretched coordinates. 

\begin{figure}
\centerline{
\epsfxsize=10.0cm \epsfbox{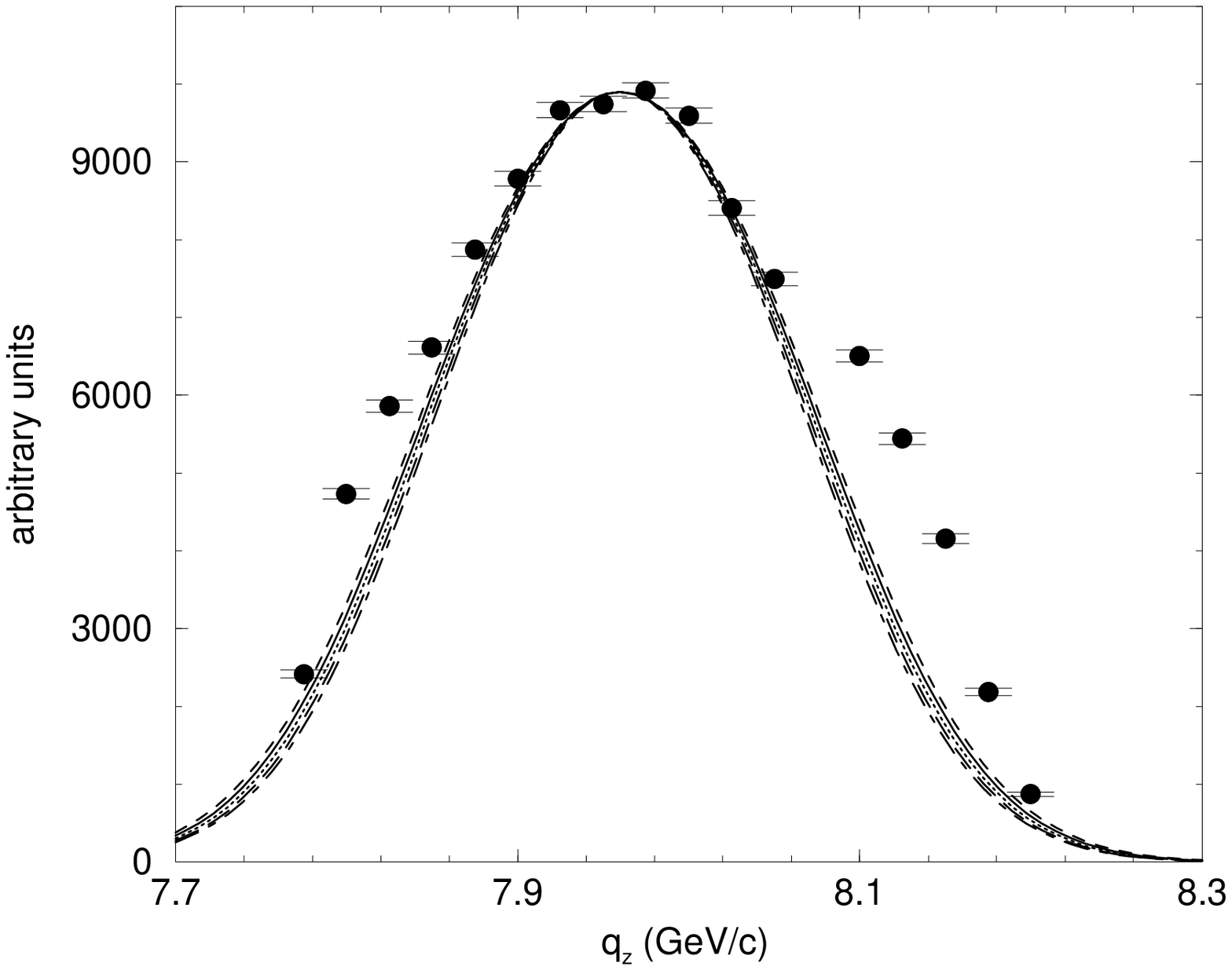}}
\caption{Longitudinal momentum distributions of $^{24}$Mg in the
stripping reaction $^{9}$Be($^{25}$Al,$^{24}$Mg)X with the $^{24}$Mg nucleus in
the intrinsic ground state. The curves correspond to different
values of the deformation parameter: $\delta$=0 (solid), $\delta$=0.1
(dotted), $\delta$=0.2 (dashed), $\delta$=0.3 (long dashed), and
$\delta$=0.4 (dash-dotted). The experimental points for the inclusive
process, together with the error bars,  from [7] are shown. The
initial beam energy is 65 MeV per nucleon.}   
\label{fff1}
\end{figure}

The longitudinal momentum distributions of the
$^{24}$Mg nucleus in the intrinsic ground state (frozen orientation)
calculated according to (\ref{frty}) for 
different values of the oscillator deformation parameter $\delta$
\cite{Boh:74,Nil:55,Nil:95} are shown in Fig.~\ref{fff1}. The 
theoretical curves are normalized to the inclusive experimental data,
the normalization constants differ for different $\delta$. We stress
again that the experimental data are inclusive and represent the sum
of differential cross sections to all final states with isospin $T$=0
and $T$=1.

Note that the case $\delta$=0 would not directly correspond to the
spherical symmetry of the projectile. Indeed, to achieve such
correspondence we have to take into account the degeneracy of
three $d_{5/2}$ sublevels with $\Omega\rightarrow 
m_{j}=\pm5/2$, $\pm3/2$, and $\pm1/2$.
In our case we move along the only Nilsson $[202\,\,5/2]$ orbital that
corresponds to a single value of $|m_{l}|=2$
\cite{Bar:98} and, of course, can not account for other
orbitals. To be consistent with the Nilsson model,
we do not draw the momentum distributions for the mixture of
different $m_j$ states.

The total stripping cross sections obtained from (\ref{frty})
by integration over $q_z$ are shown in Table~1. Although the cross
section generally  decreases with the growth of the deformation, in
the region $\delta\sim 0.3-0.5$, where the deformation parameter
corresponds to the realistic value $\delta\approx 0.34$ \cite{Lee:76}
for the $^{24}$Mg core, the cross section reaches the minimum and 
starts to increase again. This can be considered as a hint toward the
explanation of the difference between the theoretical cross section
calculated in the framework of spherical nuclei and the
experimental one. Note again that for a comparison with the case of 
sperical symmetry we need to take into account the stripping
from two additional single-particle orbitals $[211\,\,3/2]$ and
$[220\,\,1/2]$; the corresponding stripping cross sections at
$\delta=0$ are equal to 18.6 and 25.28 mb. The standard averaging
over the different orbitals gives approximately 21 mb for the cross
section corresponding to the absence of the deformation which 
agrees resonably well with the value of $\sim$26 mb, calculated for the 
spherically symmetric projectile and realistic valence proton wave
function (with correct asymptotic behavior). Our cross sections should
always be smaller than their counterparts for the functions with
correct asymptotics.

In the calculation of the longitudinal momentum distributions and
stripping cross sections we have used the following values for the  
effective target radius and minimum impact parameter: $R_T=2.948$ fm
(it corresponds to the cross section $\sigma=273$ mb for the free
proton interaction with $^{9}$Be) and $b_{min}=6.094$ fm.

\begin{center}
\vspace{1.0cm}
{\bf Table 1.} The cross sections, in mb, for the stripping of 
the $[202\,\,5/2]$ proton from  $^{25}$Al to the $^{24}$Mg ground state
rotational band at different deformation parameters $\delta$. \\
\vspace{1.0cm}
\begin{tabular}{|c|c|c|c|c|c|} \hline
$\delta$& 0 & 0.1 & 0.2 & 0.3 & 0.4 \\ \hline
$\sigma$& 19.2 & 18.6 & 18.2 & 18.1 & 18.3  \\ \hline
\end{tabular}
\end{center}
\vspace{1.0cm}

Let us now compare these results with the more elaborate calculations based
on the regular theoretical approach of Sec.~\ref{se1}. We limit
ourselves here by the consideration of the realistic $^{24}$Mg
deformation parameter $\delta\approx 0.34$ \cite{Lee:76} under the
assumption of axial symmetry. In
this approach we can calculate not only the total
stripping cross section summed over all core collective
rotational states, as was done above (Table~1), but also the
partial cross sections for the transitions into different 
rotational states of the core. If the results of the two rather
different approaches are consistent, the inclusive, with respect to
rotational states, cross section calculated
according to the prescription of Sec.~\ref{se2} has to coincide with the sum
of all partial cross sections obtained in the framework of the more
regular theoretical approach of Sec.~\ref{se1}. The cross sections
based on eq. (\ref{polop}) for the transitions into three
lowest levels of the $0^+$ rotational band of $^{24}$Mg are shown in
Table~2.

\begin{center}
\vspace{1.0cm}
{\bf Table 2.} The cross sections, in mb, for the stripping of $[202\,\,5/2]$
proton from  $^{25}$Al for the three lowest rotational states of the
core. The deformation parameter $\delta$=0.34. \\
\vspace{1.0cm}
\begin{tabular}{|c|c|c|c|c|} \hline
$L_f$& 0 & 2 & 4 & $\geq 6$\\ \hline
$\sigma$& 7.3 & 13.6 & 1.0 & negligibly small \\ \hline
\end{tabular}
\end{center}
\vspace{1.0cm}
The total cross section is equal to 22 mb that agrees 
with the inclusive cross section $\approx 18.2$ mb calculated
in the framework of the simple approach (see Table~1). The
agreement would be even closer if we would take into account the
value of the spectroscopic factor ${\cal A}^2_{\nu}$ in
(\ref{polop}). We have estimated the spectroscopic factor using the
particle-rotor model for the $^{25}$Al nucleus and obtained 
the total cross section which varies from 20.7 to 21.5 mb.

\begin{figure}
\centerline{
\epsfxsize=10.0cm \epsfbox{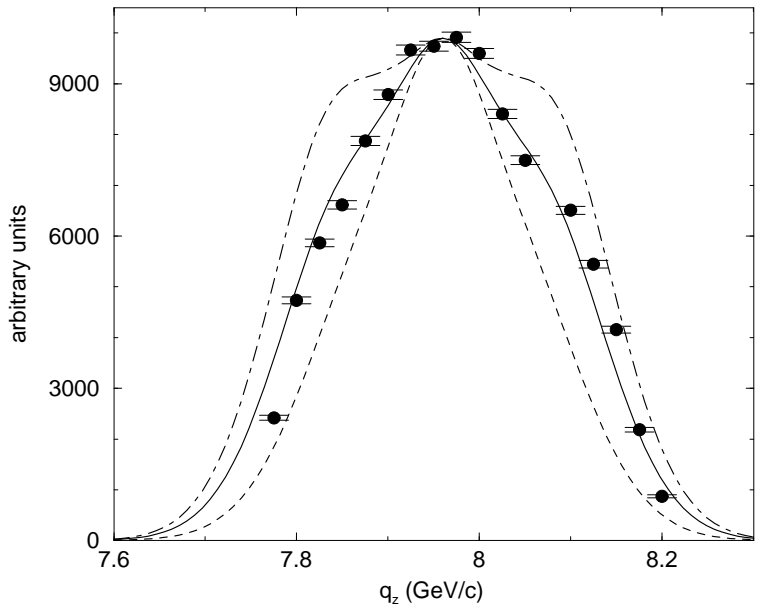}}
\caption{Longitudinal momentum distributions of $^{24}$Mg in the
stripping reaction $^{9}$Be($^{25}$Al,$^{24}$Mg$_{0^+,2^+,4^+}$)X.
The dashed curve corresponds to the transition into the $0^+$ core
state, the dashed-dotted one - to the transition into $2^+$ state,
and the solid curve is the sum of the three transitions including 
the weak $4^+$ (not shown separately). The
experimental points together with the error bars from [7] are
shown. The initial beam energy is 65 MeV per nucleon.}  
\label{fff11}
\end{figure}

As for the shape of the core longitudinal momentum distribution,
using the full calculation of Sec.~\ref{se1}
we obtain a nearly perfect agreement with experimental data as
shown in Fig.~\ref{fff11}. The dashed curve
corresponds to the transition into the $0^+$ core ground
state, the dashed-dotted one - to the transition into the $2^+$ state,
and the solid curve is the sum of the transitions to $0^+$, $2^+$,
and $4^+$ states. Note that all calculations have been performed
without any adjustable parameters except for the overall
normalization.  

\begin{figure}
\centerline{
\epsfxsize=10.0cm \epsfbox{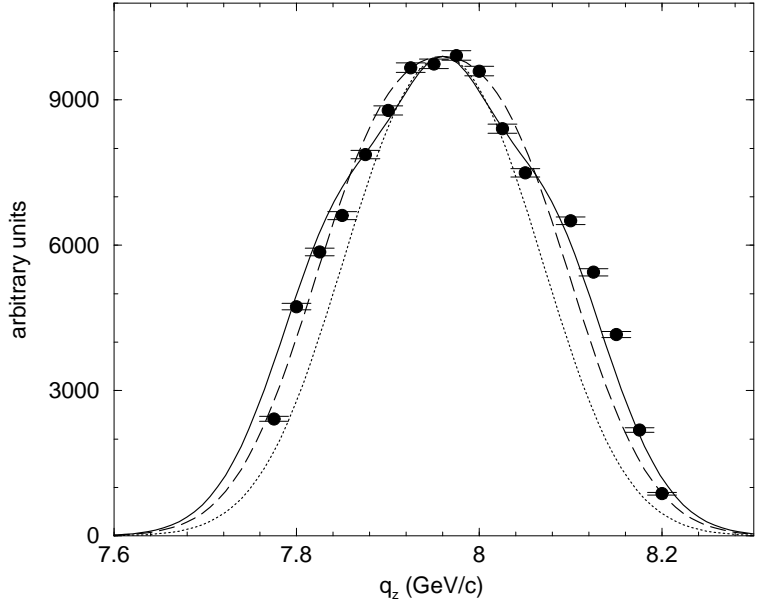}}
\caption{Longitudinal momentum distributions of $^{24}$Mg in the
stripping reaction $^{9}$Be($^{25}$Al,$^{24}$Mg)X calculated in the
framework of three different models: (i) the exact theory of Sec.~I,
solid curve; (ii) the extended geometric approach of Sec.~II, dotted
curve; (iii) the geometric approach for spherical projectiles of Ref.~[9],
dashed line. The
experimental points together with the error bars from [7] are
shown. The initial beam energy is 65 MeV per nucleon.}  
\label{ffex}
\end{figure}

Fig.~\ref{fff1} and Table~1 seemed to indicate that the deformation is
not very important both for stripping cross sections and longitudinal
momentum distributions. This agrees also with the results of
ref.~\cite{Chr:98}. However, as we see from Fig.~\ref{fff11}
and Table~2, such a conclusion would be premature. The details of the
longitudinal momentum distributions can be explained only after taking
into account the deformation of the projectile. This is clearly seen
in Fig.~\ref{ffex} where we compare the longitudinal momentum
distributions calculated in the framework of exact theory of
Sec.~\ref{se1}, the solid line, in the extended geometric approach of
Sec.~\ref{se2}, the dotted line, and in the original geometric
approach of Ref.~\cite{Han:96} developed for the spherical projectiles, 
the dashed line. The comparison of three approaches made in Fig. 4 with the
absolute normalization to the experimental peak is incomplete since the
spherical geometric approach, as a rule, leads to the cross sections of 
considerably reduced magnitude (up to 25\%).

Three features are worth pointing out in connection with the
comparison in Fig.~\ref{ffex}. The first is that although, as noted
earlier in our paper, the use of harmonic-oscillator wave functions
leads to a wrong asymptotic behavior at large distances, this problem
should be a minor one here. The reason for this is the relatively
large separation energy for the proton combined with the effect of the
Coulomb barrier, both making the tail of nucleon wave function
unimportant. Therefore the comparison with the calculation
\cite{Nav:98}, which used a $d_{5/2}$ wave function from a spherical
Woods-Saxon potential with Coulomb interaction included, can be
expected to be quantitatively meaningful. In this connection we note,
secondly, that our best approximation appears to account for the
symmetric ``shoulders'' appearing in the experimental data. We take
this as the first evidence of the necessity of incorporating the
effects of deformation in the theory of the high-energy stripping
cross sections. And, finally, as a word of caution, we remind the
reader that the experimental data represent a much more complex
situation: they are inclusive and an average over intrinsic and
rotational excitations with isospin 1 and 2 and up to 10 MeV
excitation energy. 

Below we show that there are other cases when the effects of
deformation are more significant.

\begin{figure}
\centerline{
\epsfxsize=10.0cm \epsfbox{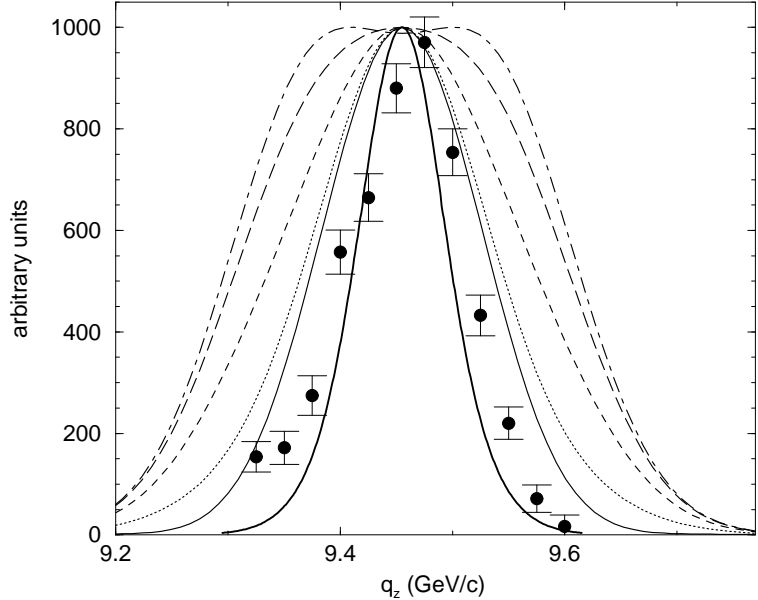}}
\caption{Longitudinal momentum distributions of $^{27}$Si in the
stripping reaction $^{9}$Be($^{28}$P,$^{27}$Si)X with the $^{27}$Si nucleus
in the intrinsic ground state. The deformation values are
$\delta =0$ for the oscillator single-particle wave function
(wide solid curve), $\delta$=0.1 (dotted),
$\delta$=0.2 (dashed), $\delta$=0.3 (long dashed), and
$\delta$=0.4 (dash-dotted). The fat solid curve corresponds to a
spherical projectile. It is calculated [7] for a Woods-Saxon
potential including the Coulomb interaction which gives the correct
asymptotics of the tail of the nucleon wave function. All theoretical
curves are normalized to the experimental data which together with
error bars are taken from ref.~[7]. The experimental data in this case
have been corrected for contributions to excited states and represent
the branch to the $^{27}$Si(5/2$^+$) ground state alone. The
corresponding partial cross section is 21$\pm$5 mb [7].
The initial beam energy is 65 MeV per nucleon. }    
\label{fff2}
\end{figure}

Let us take as a projectile a nucleus, such as $^{28}$P, which has the valence 
proton in the $[211\,\,1/2]$ Nilsson orbital. 
In reality, the $^{28}$P nucleus is spherical or very weakly deformed but
we use this example here to illustrate possible
deformation effects.
The longitudinal momentum distributions for $^{27}$Si in the
ground state calculated with eq. (\ref{frty}) are shown in
Fig.~\ref{fff2} for five values of the deformation parameter $\delta$.

\begin{figure}
\centerline{
\epsfxsize=10.0cm \epsfbox{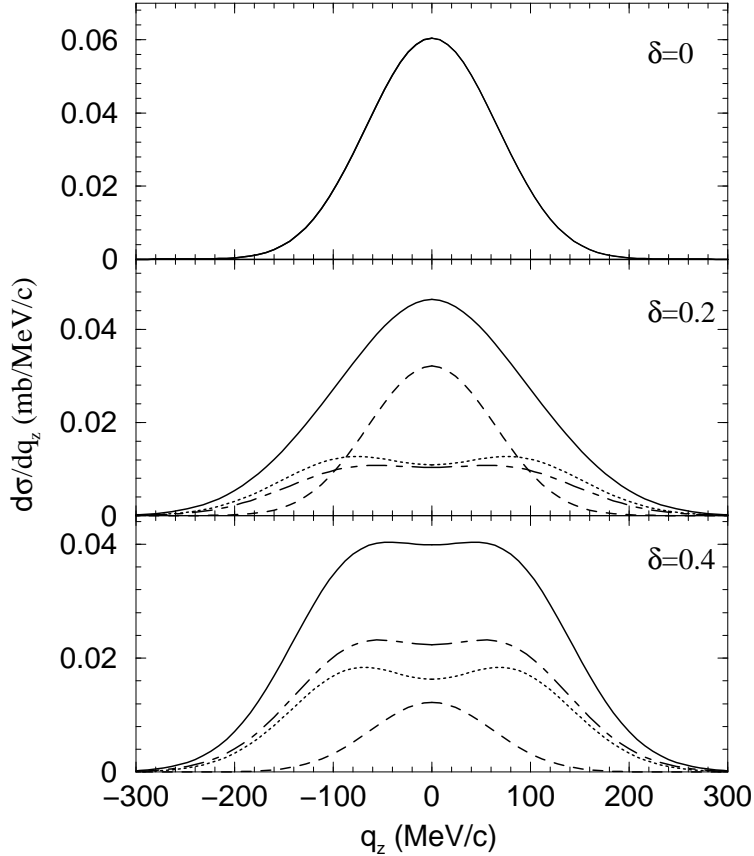}}
\caption{The total and partial longitudinal momentum distributions of
$^{27}$Si in the stripping reaction $^{9}$Be($^{28}$P,$^{27}$Si)X 
with the $^{27}$Si nucleus in the intrinsic ground state, as a
function of the assumed deformation values:
$\delta =0$ for the upper panel, $\delta$=0.2 for the middle panel,
and  $\delta$=0.4 for the lower panel. The longitudinal momentum
distributions are shown for the proton from different spherical orbitals,
$d_{5/2}$ (dotted), $s_{1/2}$ (dashed), and $d_{3/2}$ (dash-dotted);
solid curves correspond to the total distribution. At the
absence of the deformation all curves coincide (upper panel). All data
are shown in the projectile rest frame, so that the momentum scale is
different from that in Figs.~2-5.}   
\label{fff77}
\end{figure}

It is clearly seen that the width of the longitudinal momentum
distributions increases with increasing projectile deformation. The
origin of this behavior is the redistribution of the strength of
different stretched spherical components in the static field of the
deformed core, see Fig.~\ref{fff77}.  

At small deformations, Fig.~\ref{fff77}(b), the $s_{1/2}$-wave is
dominant but as the deformation increases, Fig.~\ref{fff77}(c),  the
$d_{5/2}$- and especially $d_{3/2}$-wave overcome, and as a result the
shape of the momentum distributions undergoes essential changes. It
makes it difficult (if at all possible) to determine the value of the
orbital angular momentum of the stripped nucleon from the shape of the
core longitudinal momentum distribution. At the same time these
changes allow one to  
estimate the value of the deformation parameter $\delta$. We need to bear
in mind that the oscillator hamiltonian tends to underestimate the spatial
tail of single-particle wave functions. This leads to a
wider momentum distribution compared with
asymptotically correct wave functions when we use the normalization to
the experimental data. This is
illustrated by Fig.~\ref{fff2} where the fat solid line corresponds
to the asymptotically correct spherical ($\delta$=0) wave
finction. The corresponding total stripping cross section is
close to 13 mb which is considerably less than the experimental
value $\sigma=21\pm5$ mb. The ratio of two cross sections, calculated
for the oscillator 
wave functions and for the functions with correct asymptotic behavior,
remains nearly constant in all our calculations.  
The wider solid line is related to the oscillator wave
function for the spherical case. The corresponding stripping
cross sections are shown in Table~3. 

\begin{center}
\vspace{1.0cm}
{\bf Table 3.} Cross sections (in mb) for the stripping of the
$[211\,\,1/2]$ proton from  $^{28}$P at different deformation
parameters $\delta$ calculated with harmonic-oscillator wave
functions. \\ 
\vspace{1.0cm}
\begin{tabular}{|c|c|c|c|c|c|} \hline
$\delta$& 0 & 0.1 & 0.2 & 0.3 & 0.4 \\ \hline
$\sigma$& 10.5 & 11.0 & 11.7 & 12.8 & 13.4  \\ \hline
\end{tabular}
\end{center}
\vspace{1.0cm}

\begin{figure}
\centerline{
\epsfxsize=10.0cm \epsfbox{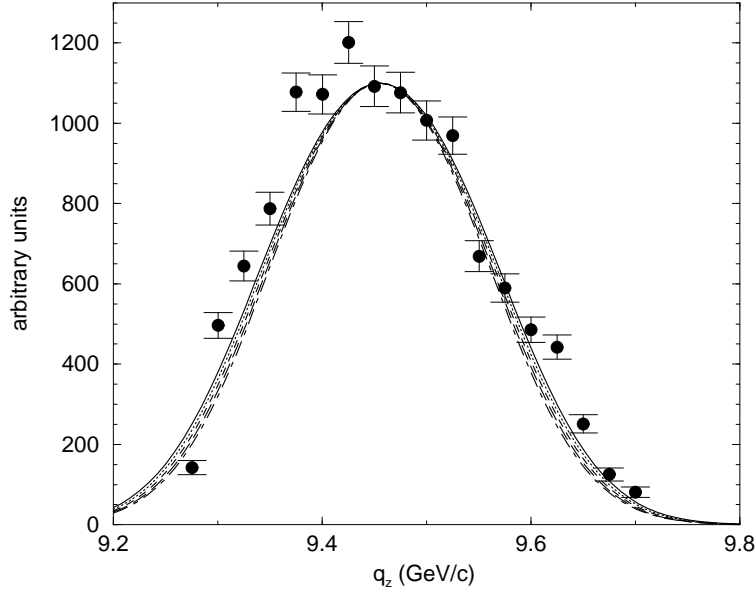}}
\caption{Longitudinal momentum distributions of $^{27}$Si 
for the hypothetical case of the
stripping reaction $^{9}$Be($^{28}$P,$^{27}$Si$^*$)X 
with the proton knocked out from the $[202\,\,5/2]$ orbital; the calculations
are based on eq. (20).
The deformation parameters are $\delta$=0 (solid curve), $\delta$=0.1 
(dotted), $\delta$=0.2 (dashed), $\delta$=0.3 (long dashed), and
$\delta$=0.4 (dash-dotted). The experimental points together with
errors from [7] are shown. The initial beam energy is 65 MeV per
nucleon.}   
\label{fff3}
\end{figure}

\begin{figure}
\centerline{
\epsfxsize=10.0cm \epsfbox{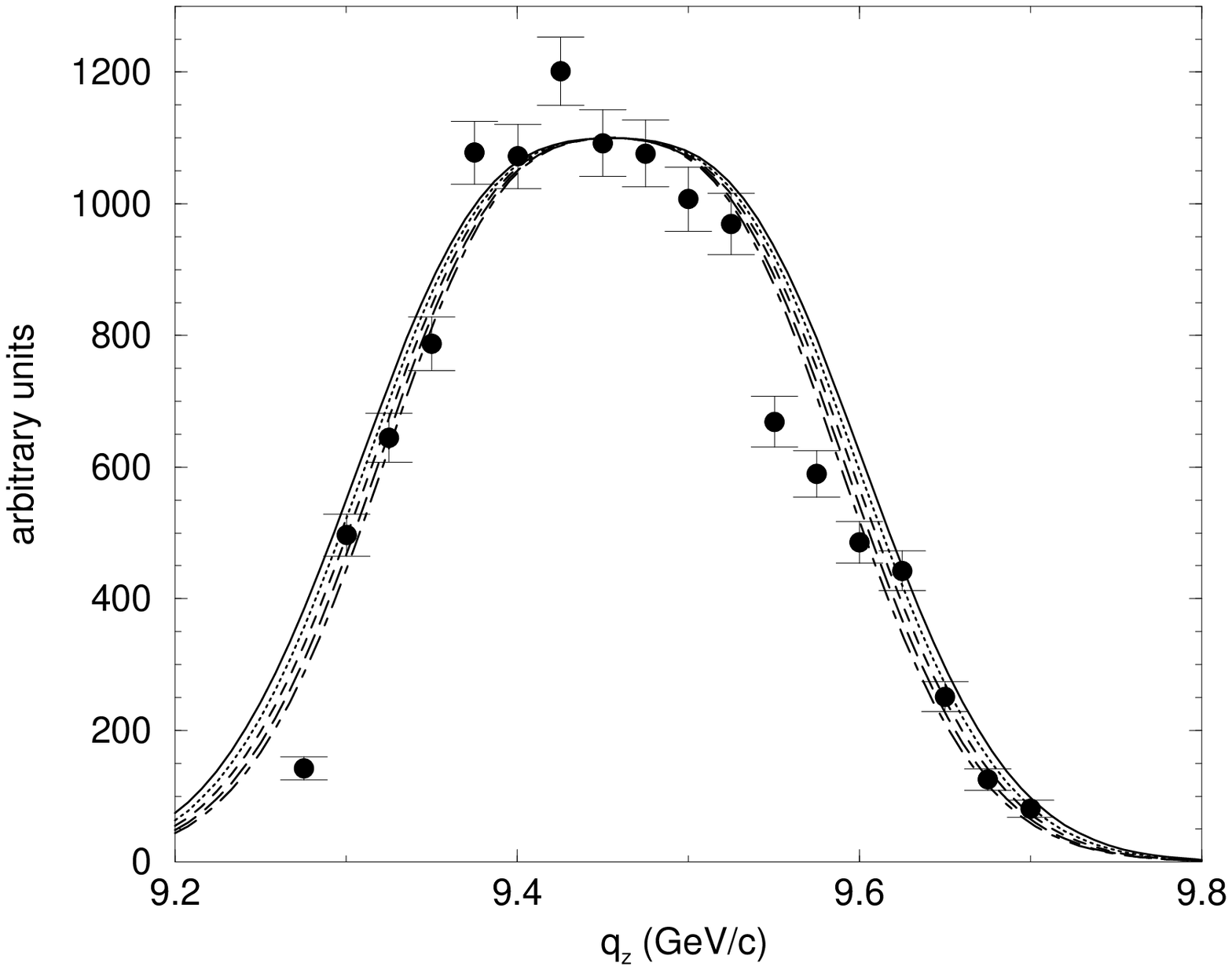}}
\caption{The same as Fig. 7, for the stripping from the $[211\,\,3/2]$
orbital.} 
\label{fff4}
\end{figure}
\begin{center}
\vspace{1.0cm}
{\bf Table 4.} The cross sections (in mb) for the stripping of $[202\,\,5/2]$,
$[211\,\,3/2]$, and $[220 \,\,1/2]$ protons from  $^{28}$P at
different deformation parameters $\delta$. \\
\vspace{1.0cm}
\begin{tabular}{|c|c|c|c|c|c|} \hline
$\delta$& 0 & 0.1 & 0.2 & 0.3 & 0.4 \\ \hline
$\sigma$ for $[202\,\,5/2]$ & 17.1 & 16.5 & 16.1 & 16.0 & 16.2  \\
\hline
$\sigma$ for $[211\,\,3/2]$ & 16.5 & 15.9 & 15.7 & 15.7 & 16.0  \\
\hline
$\sigma$ for $[220\,\,1/2]$ & 22.4 & 20.6 & 19.2 & 17.9 & 18.5  \\
\hline
$\sum \sigma$ & 55.9 & 53.0 & 52.4 & 49.7 & 50.7  \\
\hline
$\sum \sigma$ + $\sigma$ for $[211\,\,1/2]$ & 66.4 & 64.0 & 64.1 & 62.5 &
64.1  \\ 
\hline
\end{tabular}
\end{center}
\vspace{1.0cm}

\begin{figure}
\centerline{
\epsfxsize=10.0cm \epsfbox{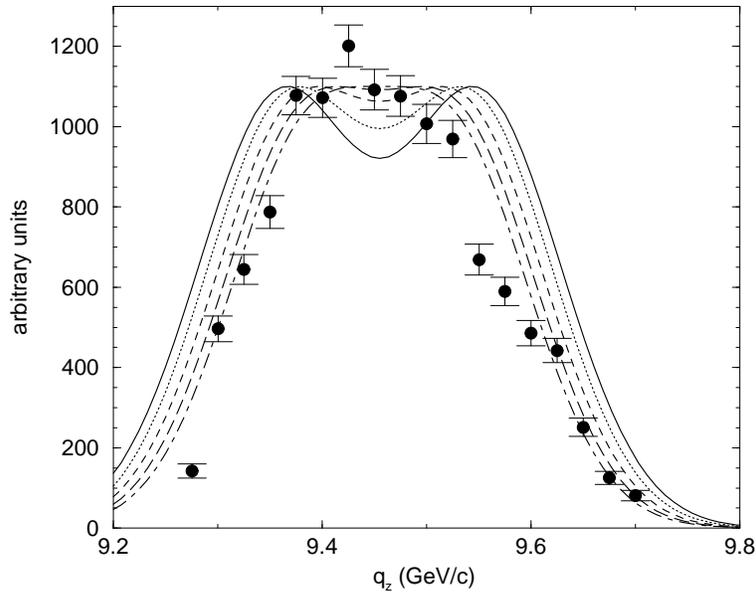}}
\caption{The same as Fig. 7, for the stripping from the $[220\,\,1/2]$ orbital.}
\label{fff5}
\end{figure}

One of the attractive features of our simple approach is the
possibility to treat the stripping reactions with the core excitation
in the same manner. For example, for the particular reaction 
$^{9}$Be($^{28}$P,$^{27}$Si$^*$)X with the detection of the excited
projectile residue $^{27}$Si$^*$, we have three orbitals in the $sd$-shell,
$[202\,\,5/2]$, $[211\,\,3/2]$, and $[220 \,\,1/2]$. 
Assuming that the measured  
\cite{Nav:98} cross section is originated by the knockout from
an individual orbital we obtain the longitudinal momentum
distributions shown in Figs.~\ref{fff3}-\ref{fff5}.
For the comparison with experimental data, $\sigma=70\pm11$ mb, we
have to sum the partial cross sections because the different core
final states were not separated in the experiment \cite{Nav:98}. The
corresponding values are shown in the last row of Table~4 including
the contribution from the ground state orbital $[211\;1/2]$ given in Table 3.
In the treatment with the use of the 
spherical shell model \cite{Nav:98} it is important to
include the relevant spectroscopic factors. In the simple Nilsson
model used here, we assume that pairing is weak and that the
occupation numbers simply take the values (0,1,2). 
We do not take into account the relevant spectroscopic factors which
depend on the deformation of the core mean field and can not be merely
borrowed from the standard shell-model calculations. To
make our consideration fully quantitative, we would need also (i) to
correct the asymptotic behavior of harmonic oscillator wave functions,
(ii) to take into account feeding from high-lying states
\cite{Han:96}. Nevertheless, even oversimplified estimations of the
partial and total 
cross sections, Table~4, show a relatively weak dependence on 
deformation whereas the shape of the longitudinal momentum
distributions, Figs.~\ref{fff3}-\ref{fff5}, is more sensitive to the
deformation parameter.   

\begin{figure}
\centerline{
\epsfxsize=10.0cm \epsfbox{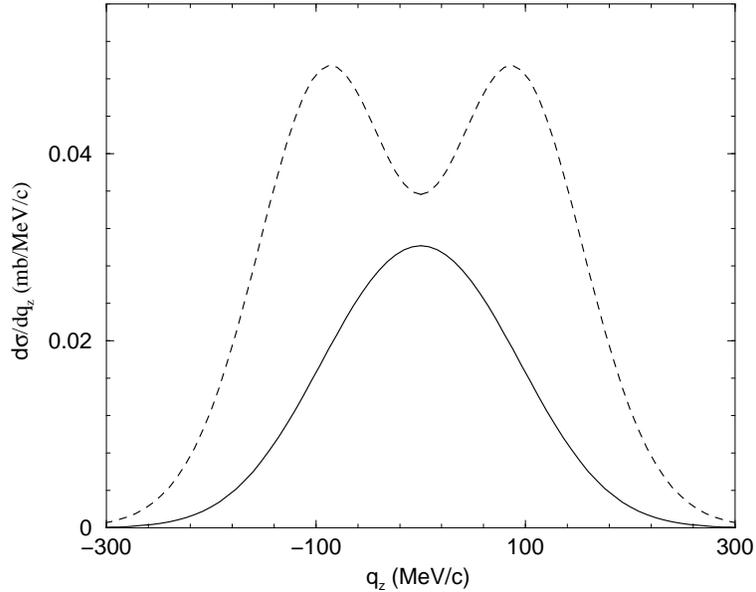}}
\caption{Longitudinal momentum distributions of $^{29}$Mg in the
stripping reaction $^{9}$Be($^{30}$Mg,$^{29}$Mg)X with the $^{29}$Mg nucleus
in the intrinsic ground state. 
The neutron is knocked out from the $[202\,\,5/2]$ orbital, solid
curve, and from the intruder $[321\,\,3/2]$, dashed curve. 
The deformation parameter for both curves is $\delta$=0.5. The
distributions are calculated in the projectile rest frame.}
\label{fff6}
\end{figure}

Based on those results we may try to predict the shape of the
longitudinal momentum distributions of the projectile residue in 
reactions with strongly deformed 
nuclei. As a possible candidate for the role of the 
deformed projectile we can use neutron-rich 
$^{26,28}$Ne or $^{30,32,34}$Mg isotopes. The measurements and
analysis of recent work \cite{Pri:99} allow one to expect
a large deformation of these nuclei. Therefore, it would be
interesting to study the longitudinal momentum distributions of the
projectile residue, for example in the reaction
$^{9}$Be($^{30}$Mg,$^{29}$Mg$_{g.s.}$)X. The preliminary
estimate based on the measurement of the reduced transition
probability $B(E2;0^+_{g.s.}\rightarrow2^+)$ gives 
$\delta\approx0.46-0.52$ for the deformation parameter of the
projectile. It corresponds to the knockout of a neutron either from 
$[321\,\,3/2]$ or from $[202\,\, 5/2]$ Nilsson orbitals (here we extrapolate
the appropriate Nilsson diagram to large
deformations). The first orbital comes out from the $pf$-shell (intruder)
whereas the second one belongs to the $sd$-shell. The orbitals intersect 
just in the region of $\delta$ indicated by the experiment \cite{Pri:99}. 
Thus, we obtain an interesting possibility to determine 
actual deformation of $^{30}$Mg via the comparison with
experimental data  and clarify
the internal structure of this nucleus. Similar arguments are true
for other isotopes in this region. To illustrate this possibility we
have calculated, see Fig.~\ref{fff6}, the longitudinal momentum distributions 
for the neutron knocked out from $[202\,\, 5/2]$ Nilsson orbital, solid curve,
and from the intruder orbital $[321\,\,3/2]$, dashed curve. 
Two results differ very strongly both
in the magnitude of the cross section and, which is even more essential, 
in shape. The
estimate of the corresponding total cross sections gives $\approx 8$ mb
for the   $[202\,\, 5/2]$ and $\approx 16$ mb for the   $[321\,\,3/2]$
orbital.

\section{CONCLUSION}
\label{se4}

We investigated the influence of the projectile
deformation on stripping cross sections and 
longitudinal momentum distributions of the core. 
For this purpose we developed an
appropriate formalism treating the deformed projectile mean field 
as in the standard Nilsson model. In practical
calculations, we limited ourselves by the fast processes 
with the core nucleus as a spectator its orientation being adiabatically fixed
during the short collision time. The study of 
reactions like $^{9}$Be($^{25}$Al,$^{24}$Mg$_{g.s}$)X, 
$^{9}$Be($^{28}$P,$^{27}$Si$_{g.s.}$)X, and
$^{9}$Be($^{28}$P,$^{27}$Si$^*$)X shows the following.\\
(i) The general influence of the deformation on the stripping cross section
and the core longitudinal momentum distribution is relatively weak
for the knockout of nucleons from orbitals which are not, or are only 
slightly, mixed by the deformed mean field of the
projectile. Nevertheless, even in such cases it is possible to
reproduce the details of the experimental longitudinal momentum
distributions only taking into account the deformation effects.\\ 
(ii) The deformation of the projectile is seen the most clearly in the
knockout of nucleons from the strongly mixed orbitals. In the case of the
axial symmetry it corresponds to the lowest possible $\Omega$. \\
(iii) The influence of deformation may be less pronounced in the case
of halo systems where the wave function of the valence nucleon is
extended far outside the core. The choice of $^{11}$Be as a projectile
\cite{Aum:99} seems to be a good testing ground for verifying this
conclusion.\\ 
(iv) The use of harmonic-oscillator wave functions with a wrong
asymptotic behavior at large distances is quite meaningful
for a projectile with high nucleon separation energy. \\
(v) The geometric approach to the description of the stripping
reactions is quite applicable for an estimation of the reaction cross
sections but it is oversimplified for reproducing the longitudinal
momentum distributions where we need to use a more detailed
semi-microscopical treatment. \\ 
(vi) The developed in Seq.~\ref{se1} semi-microscopic approach appears
to account for the symmetric ``shoulders'' of the longitudinal
momentum distributions appearing in the experimental data. We take
this as the first evidence of the necessity of incorporating the
effects of deformation in the theory of the high-energy stripping reactions. \\
(vii) The analysis of the shape of the core longitudinal momentum
distribution allows one to estimate the degree of the projectile
deformation. This is necessary for determining the 
internal structure of radioactive nuclei. The region of neutron-rich 
heavy magnesium isotopes would be a promising field for future
experiments.\\
(viii) It would be interesting to demonstrate in further calculations the
predicted by a simpler approach difference in the momentum distributions of a
strongly deformed core for the processes leading to the individual members of a
rotational band. \\
(ix) In the case of the deformed projectile it is impossible to
neglect the spin-orbit interaction since it defines the single-particle
orbitals in the deformed core mean field. \\
(x) An analysis of experimental data
ignoring the effects of the deformation  can lead to significant
errors in assigning orbital angular momentum involved in the stripping reaction.

\begin{center}
{\bf ACKNOWLEDGMENTS}
\end{center}
We are thankful to J.A. Tostevin for constructive 
discussions and reading the manuscript, and to A.~Navin for discussion
of experimental data. This work was supported by the National Science
Foundation, through Grant 96-05207.


\begin{thebibliography}{99}
\bibitem{Butler} S.T. Butler, {\sl Nuclear Stripping Reactions} (Wiley, New
York, 1957).
\bibitem{Sit:90} A.G.~Sitenko, {\sl Theory of Nuclear Reactions}
(World Scientific, Singapore, 1990).
\bibitem{Ser:47} R.~Serber, Phys. Rev. {\bf 72}, 1008 (1947).
\bibitem{Tani:95} I.~Tanihata, Prog. Part. Nucl. Phys. {\bf
35}, 505 (1995).
\bibitem{Bar:96} F.~Barranco, E.~Vigezzi, and R.A.~Broglia,
Z. Phys. {\bf A 356}, 45 (1996). 
\bibitem{Han:99} P.G.~Hansen, J. Phys. {\bf G 25}, 727 (1999).
\bibitem{Nav:98} A.~Navin, D.~Bazin, B.A.~Brown, G.~Gervais,
T.~Glasmacher, K.~Govaert, P.G.~Han- sen, M.~Hellstr\"{o}m,
R.W.~Ibbotson, V.~Maddalena, B.~Pritychenko, H.~Scheit, B.M.~Sherill,
M.~Steiner, J.A.~Tostevin, and J.~Yurkon, Phys. Rev. Lett. {\bf 81},
5089 (1998).
\bibitem{Tos:99} J.A.~Tostevin, J. Phys. {\bf G 25}, 735 (1999). 
\bibitem{Han:96} P.G.~Hansen, Phys. Rev. Lett. {\bf 77}, 1016 (1996).
\bibitem{Hen:96} K.~Hencken, G.~Bertsch, and H.~Esbensen, Phys. Rev.
{\bf C 54}, 3043 (1996).
\bibitem{Bar:98} F.~Barranco and E.~Vigezzi, In: {\sl International School of
Heavy-Ion Physics, 4th Course: Exotic Nuclei}, edited by R.A.~Broglia
and P.G.~Hansen (World Scientific, Singapore, 1998), p.~217.
\bibitem{Wild:80} B.H.~Wildenthal and W.~Chung, Phys. Rev. {\bf C 22},
2260 (1980).
\bibitem{Hey:91} K.~Heyde and J.L.~Wood, J. Phys. {\bf G 17}, 135
(1991). 
\bibitem{Cau:98} E.~Caurier, F.~Nowacki, A.~Poves, and J.~Retamosa,
Phys. Rev. {\bf C 58}, 2033 (1998).
\bibitem{War:90} E.K.~Warburton, J.A.~Becker and B.A.~Brown, 
Phys. Rev. {\bf C 41}, 1147 (1990).
\bibitem{Pri:99} B.V.~Pritychenko, T.~Glasmacher, P.D.~Cottle,
M.~Fauerbach, R.W.~Ibbotson, K.W.~Kemper, V.~Maddalena, A.~Navin,
R.~Ronningen, A.~Sakharuk, H.~Scheit, and V.G.~Zelevinsky, Phys. Lett. B,
to be published. 
\bibitem{Rid:98} D.~Ridikas, M.H.~Smedberg, J.S.~Vaagen, and
M.V.~Zhukov, Nucl. Phys. {\bf A 628}, 363 (1998).
\bibitem{Chr:98} J.A.~Christley and J.A.~Tostevin, Phys. Rev. {\bf C
59}, 2309 (1999).
\bibitem{Boh:74} A.~Bohr and B.R.~Mottelson, {\sl Nuclear Structure},
Vol. 2 (Benjamin Inc., New York, 1974).
\bibitem{Dro1:55} S.I.~Drozdov, Sov. Phys. JETP {\bf 1}, 588, 591 (1955); 
{\bf 7}, 889 (1958).
\bibitem{Ino:56} E.V.~Inopin, Sov. Phys. JETP {\bf 3}, 134 (1956).
\bibitem{Abg:79} Y.~Abgrall, J.~Labarsouque, and B.~Morand, Nucl. Phys. {\bf
A316}, 389 (1979).
\bibitem{Fal:90} G.~F\"{a}ldt and R.~Glauber, Phys. Rev. {\bf C
42}, 395 (1990).
\bibitem{Fal:80} G.~F\"{a}ldt, J. Phys. {\bf G 6}, 1513 (1980).
\bibitem{Nis:82} H.~Nishioka and R.C.~Johnson, J. Phys. {\bf G 8}, 39
(1982). 
\bibitem{Yab:92} K.~Yabana, Y.~Ogawa, and Y.~Suzuki, Nucl. Phys. {\bf
A 539}, 295 (1992).
\bibitem{Fuj:80} T.~Fujita and J.~H\"{u}fner, Nucl. Phys. {\bf
A343}, 493 (1980). 
\bibitem{AlK:95} J.S.~Al-Khalili, I.J.~Thompson, and J.A.~Tostevin, 
Nucl. Phys. {\bf A 581}, 331 (1995).
\bibitem{Gla:59} R.J.~Glauber, In: {\sl Lectures in Theoretical
Physics}, edited by W.E.~Brittin and L.G.~Dunham (Interscience, New
York, 1959), Vol. I, p.~315.
\bibitem{Gla:70} R.J.~Glauber and G.~Matthiae, Nucl. Phys. {\bf B21},
135 (1970).
\bibitem{Sit:57} A.I.~Akhieser and A.G.~Sitenko, Phys. Rev. {\bf 106},
1236 (1957).
\bibitem{Hus:85} M.S.~Hussein and K.W.~McVoy, Nucl. Phys. {\bf A445},
124 (1985).
\bibitem{Hus:89} M.S.~Hussein and R.C.~Mastroleo, Nucl. Phys. {\bf A491},
468 (1989).
\bibitem{Ban:93} P.~Banerjee and R.~Shyam, Nucl. Phys. {\bf A 561},
112 (1993).
\bibitem{Boh:69} A.~Bohr and B.R.~Mottelson, {\sl Nuclear Structure},
Vol. 1 (Benjamin Inc., New York, 1969).
\bibitem{War:92} E.K.~Warburton and B.A.~Brown, Phys. Rev. {\bf C 46},
923 (1992).
\bibitem{Joa:} J.~Joachain, {\sl Quantum Collision Theory}
(North-Holland, Amsterdam, 1980).
\bibitem{Nil:55} S.G.~Nilsson, Kgl. Danske Vidensk. Selsk.
Mat.-Fys. Medd. {\bf 29}, No. 1 (1955). 
\bibitem{Nil:95} S.G.~Nilsson and I.~Ragnarsson, {\sl Shapes and
Shells in Nuclear Structure} (Cambridge University Press, Cambridge,
1995). 
\bibitem{Edm:60} A.R.~Edmonds, {\sl Angular Momentum in Quantum
Mechanics} (Princeton University Press, Princeton, 1960).
\bibitem{Lee:76} E.W.~Lees, C.S.~Curran, T.E.~Drake, W.A.~Gillespie,
A.~Johnston, and R.P.~Singhal, J. Phys. {\bf G 2}, 105, (1976).
\bibitem{Aum:99} T.~Aumann {\sl et al.}, to be published.

\end{thebibliography}
\end{document}